\documentstyle[twocolumn,prl,aps]{revtex}

\begin{document}

\draft
\twocolumn[\hsize\textwidth\columnwidth\hsize \csname @twocolumnfalse\endcsname

\title{On the singular spectrum of the Almost Mathieu operator.\\
Arithmetics and Cantor spectra of integrable models.}

\author{P. B. Wiegmann}
\address{James Franck Institute
and Enrico Fermi Institute
of the University of Chicago,
5640 South Ellis Avenue, \\
Chicago, Illinois 60637 \\
and
Landau Institute for Theoretical Physics}

\maketitle

\begin{abstract}
I review a recent progress towards solution of the Almost Mathieu
equation (A.G. Abanov, J.C. Talstra, P.B. Wiegmann, 
Nucl. Phys.  B 525, 571 ,1998), known also as Harper's equation or Azbel-Hofstadter
problem.
 The
spectrum of this
 equation is known to be a pure singular continuum with a rich hierarchical structure.
Few years ago it has been found that the almost Mathieu operator is integrable. An
asymptotic solution of this operator  became
possible due analysis the Bethe Ansatz equations.
\end{abstract}

 ]

1. {\it Introduction} In this lecture I review a recent progress \cite{ATW97} towards
decoding of one of the most puzzling strange set generated by a quasiperiodic Schr\"odinger
operator :
\begin{equation}
\label{harper}
\psi_{n+1}+\psi_{n-1}
+2\lambda\cos(\theta+2\pi n\eta)\psi_n=E\psi_n \end{equation}
The history of this equation as well as its applications in different branches of
physics and mathematics are rich. This equation, known as Harper's equation,
describes the electronic spectrum of one-dimensional quasicrystal (a particle in a
quasiperiodic potential) and often used to study localization-delocalization
transition (see e.g. \cite{AubryAndre80,Sinai87}). It also describes a Bloch
particle in a uniform magnetic field and also known as 
Azbel-Hofstadter problem \cite{Azbel64,Hofstadter76}. It is a standard example
of the operator, also known as {\it almost Mathieu operator} with a singular
continuum spectrum \cite{BelSim82,LastJit94}. This list may be continued.

The spectrum of this equation is complex. In the commensurate case, when 
$\eta$ is a rational number
$$\eta=P/Q,$$
one may impose the Bloch condition:
$$\psi_n=e^{ik'}\psi_{n+Q}$$
Then, 
the spectrum consists of
$Q$ bands, separated by $Q-1$ gaps. In the incommensurate limit, when  $\eta$ is an irrational
number\footnote{Although some properties of the spectrum depend on the 
type of irrational number $\eta$, here we consider  
{\it typically Diophantine} numbers. These numbers have a full
Lebesgue measure and thus sufficient for almost all physical
applications.}  ($P\rightarrow
\infty,\,\,\,Q\rightarrow\infty$)  the spectrum becomes an {\it infinite Cantor
set}\footnote{closed,  nowhere dense set with no isolated
points}\cite{Azbel64,BelSim82}   with total bandwidth (Lebesgue measure of the
spectrum) $4|\lambda -1|$
\cite{AubryAndre80,ThoulessBW83,AMS90}.  

The most interesting  ``critical'' case appears at $|\lambda| =1$. Then the spectrum 
becomes a {\it purely singular
continuum} \cite{LastJit94}.
 In this case  wave functions lost their extended character and not yet localized
but exhibit  a power law scaling.  Moreover, there
is  numerical evidence and almost a consensus, that in this
case ($\lambda=1$) the spectrum and wave functions are multifractal \cite{TangKohmoto83}.

Multifractal sets exhibit a sort of conformal invariance and are expected to be described by
methods of conformal field theory. This theory, however, is yet to be developed and scaling
properties of sets generated by dynamical systems and by closely related quasiperiodic
systems are far from being understood. 

Since the empirical observations of Hofstadter \cite{Hofstadter76}, the evidence has
been mounting that the spectrum of (1) ({\it Hofstadter butterfly}) as well as other
quasiperiodic equations with a differential potentials are regular and universal rather
than erratic or ``chaotic''. Few years ago it has been
shown\cite{WiegmannZab94,FadKash95,Kutz,KreftSeiler96,F} that despite the complexity of the
spectrum, the Harper-Azbel-Hofstadter-almost Mathieu operator
(\ref{harper}) equation at any rational
$\eta=P/Q$ is integrable and can be ``solved'' by employing methods of integrable systems
Ansatz (BA). This had opened the possibility of describing the complex behavior of
an incommensurate system as a limit of a sequence of integrable models. I hope that this solution
will help to apply  conformal field theory to dynamical systems.

The symmetries of the problem which eventually lead to its integrability are the most
transparent it its "magnetic" interpretation. Consider a particle on a two dimensional square
lattice in a magnetic field with a flux $\Phi=2\pi\eta$ per plaquette. Its Hamiltonian is:
$$H=T_x+T_x^{-1}+\lambda(T_y+T_y^{-1}),$$
where operators $T_x$ and $T_y$ describe translations of the particle
 in $x$ and $y$ direction by a lattice site. In magnetic field translations form a Weyl pair:
\begin{equation}%
 T_xT_y=q^2T_yT_x,$$    
 \label{TT}
\end{equation}%
where
$$q^2=e^{i\Phi}.$$

The Harper's equation, then appears as a result of representation of translation operators as
a shift and multiplication:
\begin{eqnarray}%
  &&T_x\psi_n=\psi_{n+1},\nonumber\\
&&T_y\psi_n=q^2e^{ik}\psi_n   
 \label{TTT}
\end{eqnarray}%

2. {\it Hierarchical tree.} I begin by describing the scaling hypothesis for the spectrum of
an incommensurate (quasiperiodic) operator with a purely singular continuum spectrum. To do
so, we need a notion of the  hierarchical tree.

Let us consider a sequence of rational
approximants $\eta^{(j)}=P_j/Q_j$ with increasing $Q_j$ 
to an irrational flux
$\eta$ so that:
$|\eta^{(j)}-\eta|<c(Q_j)^{-2}$, where $c$ is  a
$j$-independent constant\footnote{This sequence always exists and
can be constructed e.g.\ from the Farey Series\cite{Vinogradov55}}. A
Harper equation taken for each $\eta^{(j)}$ is generations of the hierarchy. Let us consider a
graph (with no loop), which connects the $k$-th band of the generation
$\eta^{j}$ (the daughter generation) to a certain  band
$k'$ of some previous (parent) generation $\eta^{(j-1)}$.
 We call it a {\it hierarchical tree}
if energies
$E_j({\cal J})$ belonging to any branch ${\cal J}$ of the tree form a 
sequence converging to the point
$E({\cal J})$ of the spectrum in such a way that the sequence
$ Q^{2-\epsilon_{\cal J}}|E_j({\cal J})-E({\cal J})|$ is bounded but 
does not converge to zero. 
 
The set of numbers $\epsilon_{\cal J}$ are anomalous 
exponents. In a multifractal spectrum, anomalous dimensions depend on the branch. They and
the tree characterize ultrametric properties of the spectrum.

Let us stress that the very existence of the hierarchical tree is a hypothesis and the tree
constructed below is the conjecture. We call it {\it scaling hypothesis}.

To
construct the hierarchical tree it is necessary to find the sequence of
generations and a rule to determine the parent generation and a parent band out of a given
band of a given generation. In other words, the hierarchical tree is determined by 
 a sequence of rational approximants $\eta^{(j)}\rightarrow \eta$ and by a
mapping 

\begin{equation}%
  (k,P,Q)\rightarrow (k',P',Q').   
 \label{map}
\end{equation}%
where $k$ and $k'$ are labels of a daughter's band  and the parent's band of generations
$P/Q$ and $P'/Q$ respectively. To describe the hierarchical tree we will need a notion of a
{\it discrete spectral flow} and its rate {\it Hall conductance}. 

A heuristic definition of the spectral flow is as follows. Let us consider a spectrum of the 
problem with a given $\eta$ and choose some (big) gap. We label it by $k$. Now let us change
$\eta$ by a small $\delta \eta$, such that newly appeared gaps in the vicinity of the edge
of the big gap, will be smaller than the big gap $k$. Then we can look on new levels appeared
within the big gap, close to the bottom of the gap. 
  The number of these levels,  i.e., the number
of levels $\delta N_k$, crossing an energy
$E$, lying inside the gap, close to its lower edge, is a spectral flow. One can also treat the
spectral flow as a number of levels appeared within a "big" band adjacent to the gap from
below. The rate of the spectral flow is
\begin{equation}
\sigma_k = \frac{\delta N_{k}}{\delta\eta}.
\label{StredaF}
\end{equation}
is known to be the Hall conductance of the gap \cite{Streda} (Streda's formula) of two
dimensional particles in magnetic field. This number
 is an integer and depends on the gap. The difference between the Hall
conductances of  nearest gaps, i.e., the spectral flow into the k-th band
\begin{equation}
\sigma(k)=
\sigma_{k}-\sigma_{k+1}
\label{1}
\end{equation}%
 is the Hall conductance of the band. 

The index theorem identifies  the Hall conductance of the band with the number of zeros of
the wave function $\psi_n(k,k')$ within the Brillouin zone: $0<k<2\pi,\,0<k<2\pi/Q$,
i.e., with the Chern class of a band:
$$\sigma(k)=\frac{1}{2\pi}\oint_{\partial B}\vec\nabla\ln \psi_n(k,k')d\vec k,$$
where the integral goes over the boundary of the magnetic Brillouin zone of the $k$ th band. 

The Hall conductivity of the k-th gap varies the
range $-Q/2<\sigma_k\leq Q/2$ and 
obeys the Diophantine 
equation\cite{TKNN82,DAZ85}
\begin{equation}
\label{DE}
P \sigma_k = k\; (\mbox{mod}\; Q).
\end{equation}
In it turns the Hall conductance of a band $\sigma(k)$ is allowed to  have only two values.
They can be found
 explicitly. Let us consider a continued fraction expansion of
\begin{equation}
   \eta^{(j)}=
    \frac{1}{n_1+\frac{1}{n_2
    +\frac{1}{n_3+\ldots}}}
    \equiv [n_1,\,n_2,n_3,\ldots,\,n_j ]
  \label{parappr}
\end{equation}
Then the Hall conductance of the gap $k$ is 
\begin{equation}
\label{HC}
{\sigma_k} = \frac{Q_j}{2}-Q_j\left\{(-1)^j  \frac{Q_{j-1}}{Q_j}k
+\frac{1}{2}\right\},
\end{equation}
while the two values of the Hall conductance of the band $k$ may be
\begin{equation}%
  \sigma(k)=
(-1)^{j-1}Q_{j-1},\;\;\; \mbox{or}\;\;\;
(-1)^{j}(Q_j-Q_{j-1})
\label{Hall}
\end{equation}%
here $\{x\}$ is fractional part of $x$.

Now let us turn to the hierarchical tree. We conjecture that the hierarchical tree is the
spectral hierarchy - an integral version of the spectral flow.
 Let us  consider two close rational  $P/Q$ and
$P'/Q'$ with
$Q'<Q$. The number of states per lattice site in a band is $1/Q$ and $1/Q'$ respectively. If
there is a band $k$ of the problem with a flux $P/Q$, such that its Hall
conductance is a ratio between the difference of the number of states and the
fluxes
\begin{equation}\label{integrstreda}%
\frac{1}{Q}-\frac{1}{Q'}=\sigma(k) (\frac{P}{Q}-\frac{P'}{Q'})
\end{equation}%
then we say that the band $k$ of the generation $P/Q$ has a "parent"
band in the generation $P'/Q'$. The absolute value of the Hall conductance is the difference
between  number of states in the parent and a daughter band:
\begin{equation}
Q-Q'=|\sigma(k)|
\label{zx}
\end{equation}%

This formula may be viewed as an integrated Streda formula  (\ref{StredaF}).
It determines the flux $P'/Q'$ and and by virtue of an iterative procedure,
generates a sequence of rational approximants (generations of the hierarchical tree),
$\eta^{(j)}$ to
 $\eta$.

The integrated Streda formula is not enough to determine the tree.  We complete it by the {\it
adiabatic} principle, which states that the levels do not cross each other along a tree. This
proposition may be put in symbols. Let us enumerate all states from the bottom of the spectrum
and characterize them by a fraction
$\nu=\mbox{(number of the state)}/Q$. All states in the k-th band have the
fraction $(k-1)/Q<\nu<k/Q$. The adiabatic principal assumes that a state in
the middle of the band $k$ with a fraction $(k-1/2)/Q$ has a parent somewhere in the
parent band $k'$, i.e.\ 
\begin{equation}\label{k}
(k'-1)/Q'<(k-1/2)/Q<k'/Q'.
\end{equation}
This  together with integrated Streda  formula (\ref{integrstreda}) and formulas for
the Hall conductance (\ref{1},\ref{DE},\ref{Hall}) determines the mapping
(\ref{1}) and thus the hierarchical tree. For quadratic irrationals this algorithm
allows to find the tree analytically. I do not concentrate here on this matter, but would
like to make two comments:

{\it Intermediate fractions.} Rational approximants (generations of the tree), generated by
this procedure, appeared to be differ from truncated continues fraction of $\eta$. They are:
\begin{equation}\label{323}
\eta^{(j-1)}=\left\{
\begin{array}{ll}
\left [n_1,\ldots,n_j-1\right ],&
if\,\,\sigma(k)=(-1)^{j-1}Q_{j-1}\\
\left [n_1,\ldots,n_{j-1}\right ],&
if\,\sigma(k)=(-1)^{j}(Q_{j}-Q_{j-1})
\end{array}\right.
\end{equation}
Thus the parent generation is obtained by either subtracting 1 from the last
quotient $n_j$ of the continued fraction or by truncating the fraction. The
sequence of generations produced by these iterations is known as
 intermediate fractions. The golden mean $\eta=\frac{\sqrt{ 5}-1}{2}$ is the only number
which rational approximants are truncated continues fraction.

{\it Takahashi-Suzuki numbers}. The entire set of Hall conductances generated by
the flux $P/Q$ are numbers less than $Q$
of the form
\begin{equation}
   \label{TZ}
   \{Q_{i-2}+mQ_{i-1}\;|\;m=1,\ldots,n_i,\;\;i=1,2,\ldots\}
\end{equation}

These numbers (denominators of intermediate fractions) are known in
integrable models related to $U_q(SL_2)$ as Takahashi-Suzuki numbers
\cite{TakSuz72}.  They are allowed lengths of string solutions of Bethe-Ansatz equations. 
These numbers are also a set of possible dimensions of irreducible highest weight 
representations of
$U_q(SL_2)$  with definite parity
\cite{MezNep90}.

Each path of the tree may also be characterized by a fraction $\nu^{(j)}_k=k/Q_j$ lying
on the path and converged to a given irrational fraction $\nu$. According to eq.(\ref{k}) the
parent fraction is determined by the daughter one as 
$\nu^{(j-1)}=\frac{1}{Q_{j-1}}\left(\left[Q_{j-1}(\nu^{(j)}
-\frac{1}{2Q_{j}})\right]+1\right)$ with $Q_{j-1}=Q_j-|\sigma(k)|$. 
The sequence $\nu^{(j)}$ converges to the irrational $\nu$ faster than 
$Q_j^{-1}$, i.e.,  $|\nu^{(j)}-\nu |<c Q_j^{-1}$. In
terms of the fractions one may reformulate the scaling hypothesis as
$|E_j-E|<c|\nu^{(j)}-\nu |^{\alpha({\cal J})}$, defining the scaling exponent 
$\alpha({\cal J})$. 

The hierarchical tree, we just  described, has been suggested by the Bethe Ansatz equations
for  the Harper's equation. However, it seems plausible that the construction is
universal and valid for a general quasiperiodic equation, regardless, whether it is integrable
or not. A set of Hall conductances is the only input of the algorithm.

3. {\it Integrability.}  The Harper's equation (\ref{harper}) is integrable as soon as $\eta$
is a rational. Here I adopt a restricted definition of the integrability of a linear
eqaution: there is an isospectral transformation which turns all Bloch solutions of the
Harper's equation to discrete polynomials of degree
$Q$. In symbols
\begin{equation}
\label{psi}%
\psi_n=e^{ik'n}\sum_{m=0}^{Q-1} c_{nm}\Psi_m, \end{equation}%
where $c_{nm}$ is a unitary $Q\times Q$
matrix and $\Psi(z)$ is a polynomial of the degree $Q-1$. 
In other words, there is a gauge (a choice of the gage potential), or a representation of the
algebra of translations in a magnetic field (2), where all wave functions are polynomials.

 In
this sence the Harper's equation appears to be integrable  for any point of the Brillouin zone
$0\leq k'<2\pi/Q,\;0\leq k<2\pi$ 
\cite{FadKash95},  although the Bethe Ansatz equations  look especially simple at
the so called {\it rational} points of the Brillouin zone.
The latter correspond to the centers and edges of bands.
The study of these points is sufficient for our purposes.
In this case 
$$\Psi_n=\sum_{j=0}^{Q-1}a_j(\rho q^{n})^j$$
where   $a_j$ do not depend on $n$ and $\rho$ is a constant.

 It appears to be convenient to
parameterized polynomials  by its roots:
\begin{equation}
\label{polform}
\Psi(z)\equiv\sum_{j=0}^{Q-1}a_jz^j=\prod_{i=0}^{Q-1} (z-z_i).
\end{equation}

 Below we
sketch the results of the Bethe-Ansatz solution and skip all aspects of integrability related to cyclic
representations of
$U_q(Sl_2)$ \cite{comment}. 

Rational points form a zoo. To characterize them we introduce parameters
$\tau,\,\kappa,\,\mu=\pm 1$.  The choice
$\tau=1$ yields  levels at the center of bands, while $\tau=-1$ corresponds to  edges of
bands. The Chambers relation  
\begin{equation}
   \Lambda(k',k)\equiv\mbox{det} H= 2\cos{Qk'}+2\lambda\cos{Qk}
 \label{Lambda}
\end{equation}
implies that the  energy depends on $k'$ 
and $k$  via $\Lambda(k',k)$.
Therefore the edges of the energy bands are given by extrema of $\Lambda$
which assumes a minimum/maximum given by $\Lambda=\pm (2+2\lambda)$. The middle of bands
corresponds to 
$\Lambda=0$. 
If $P$ is even ($Q$ is odd) the rational points are labeled by additional discrete parameters
$\kappa,\mu=\pm 1$. The middle points of bands at $\kappa=\pm
1$ are given by the equation $\frac{\cos\frac{Q}{2}(k_x+\pi\frac{P}{2Q})}
{\sin\frac{Q}{2}(k_y+\pi\frac{P}{2Q})} =\nu(-1)^{\frac{Q-1}{2}}$.
The edges ($\tau=-1$) of bands $k'=\frac{\pi}{Q}\frac{1-(-1)^{\frac{P}{2}}}{2}$ and
$k=\frac{\pi}{Q}\left(\frac{1-(-1)^{\frac{P}{2}}}{2}+2l\right)$ are distinguished by
parameter $\mu$. Being count from the bottom of
the spectrum, these edges are ordered as bottom-top-bottom$\ldots$ if
$\mu=-(-1)^{\frac{P}{2}}$-odd and top-bottom-top$\ldots$ if
$\mu=(-1)^{\frac{P}{2}}$ (see \cite{ATW97}) for detailes).

For the rational points the transformation (\ref{psi}) is  given by ``quantum
dilogarithms''
\begin{equation}%
c_{nm}=\prod_{j=0}^{m-1}\Big(e^{ik}q^{2n+1/2} \frac{\lambda^{1/2}+\tau\kappa\rho^{-1}
q^{-j-1/2}}{\lambda^{1/2}+\kappa\rho q^{j+1/2}}\Big), \end{equation}%
where $\rho=i\exp(i\frac{k_x+k_y-\pi P}{2})$.
Under this transformation the Harper's equation becomes:
\begin{eqnarray}
\label{HamChiral223}
&& iq \Big(z^{\frac{1}{2}} +i\tau\kappa
(\frac{1}{\lambda qz})^{\frac{1}{2}}\Big) \Big(z^{\frac{1}{2}} -i\kappa
(\frac{\lambda}{qz})^{\frac{1}{2}}\Big) {\Psi}(qz) \nonumber \\
&&-iq^{-1} \Big(z^{\frac{1}{2}} -i\tau\kappa
(\frac{q}{\lambda z})^{\frac{1}{2}}\Big) \Big(z^{\frac{1}{2}} +i\kappa
(\frac{\lambda q}{z})^{\frac{1}{2}}\Big) {\Psi}(zq^{-1})=\nonumber \\
&&\mu\kappa \lambda^{-1/2} E{\Psi}(z),
\end{eqnarray}
where one suppose to set $z=\rho q^n$. However, there is a certain advantage to
consider the difference equation for $\Psi(z)$ in a complex
plane $z$.

The integrability now reads: all Bloch solutions of the difference equation
(\ref{HamChiral223}) are polynomials.

I try to unmasked this transformation by a  comment bellow, however it becomes meaningful in
the
$U_q(sl_2)$ setup. 

 Let us represent translation operators (\ref{TTT}) by another Weyl pair
\begin{equation}%
  UV=qVU,   
 \label{uv}
\end{equation}%
and setting
\begin{equation}
 \label{bc}
	T_x=UV\frac{U+a}{U+b},\;T_y=VU^{-1}\frac{U+a}{U+b},
	\end{equation}%
where
\begin{eqnarray}
a &=& -i\nu  q^{-\frac{1}{2}} \lambda^{-1/2}\\
b &=& -i\tau\nu  q^{-\frac{1}{2}}\lambda^{1/2} 
\end{eqnarray}
Equation (\ref{HamChiral223}) appears, by choosing a standard representation  of 
$U$ and $V$:
$$(U\Psi)_n= -\rho^{-1}q^{-n}\Psi_n,\;\;
  (V\Psi)_n=-i\tau\nu \mu\Psi_{n+1}.$$

4. {\it The Bethe Ansatz.} Being sure that solutions of the Eq.(\ref{HamChiral223}) are
polynomials, we may evaluate it at one of the roots of the polynomial $z_i$. This gives the
Bethe-Ansatz (BA) equations: 
\begin{equation} 
q^{Q}\prod_{k=1}^{Q-1}\frac{qz_i-z_k}{z_i-qz_k} = \frac{\left(z_i-i\tau
\kappa \lambda^{-\frac{1}{2}}q^{\frac{1}{2}}\right) 
\left(z_i +i\kappa \lambda^{\frac{1}{2}}q^{\frac{1}{2}}\right)} {\left(q^{\frac{1}{2}}z_i
+i\tau\kappa\lambda^{-\frac{1}{2}}\right) \left(q^{\frac{1}{2}}z_i
-i\kappa\lambda^{\frac{1}{2}}\right)}.
\label{BetheAnsatz}
\end{equation}
Solutions of the BA equations give the wave functions of the 
Harper equation at band's edges and centers. Their energy is given by 
\begin{equation}
\label{en1}
E=i\mu \lambda^{\frac{1}{2}}q^{Q}(q-q^{-1})
\left[\kappa\sum_{i=1}^{Q-1}z_{i} -i\frac{\lambda^{\frac{1}{2}}
-\tau\lambda^{-\frac{1}{2}}}{q^{1/2}-q^{-1/2}}
\right]. \end{equation} The latter is obtained by evaluating the leading terms at
$z\rightarrow\infty$ at eq.(\ref{HamChiral223}). 

At first glance, the BA equations (\ref{BetheAnsatz}) look even more
complicated than the original Harper equation. This is true as long as $Q$ is not large.
However, the BA equations (\ref{BetheAnsatz}) provide a better description of the problem in
the most interesting, incommensurate, limit
$P, Q\rightarrow \infty, \eta\rightarrow\mbox{irrational number}$.

At $P$ odd the BA equations admit an exact zero mode solution at for $E=0$ \cite{HKW94}.
For a quasiclassical analysis of the BA equations at $\eta\rightarrow 0$,  see\cite{K}.

Below, we consider the most interesting case $\lambda=1$.

3. {\it String hypothesis}.
Here we formulate the string hypothesis which allows us to obtain the 
solutions of the BA equations (at $\lambda=1$) with an accuracy ${\cal O}(Q^{-2})$.
The hypothesis is based on the analysis of singularities of the BA, 
and is supported by extensive numerics \cite{ATW97}. 
 Here we just formulate the string hypothesis and present
some immediate consequences. To proceed, we
need the notion of {\it strings}.

 A string of spin $l$, parity $v_=\pm 1$ and center $x_l$ is a set 
of $2l+1$ complex
numbers:
\begin{equation}
 \label{String2Def}
    z_m^{(l)}=v_lx_l
   q_{l}^m,\,\,\,\,m=-l,-l+1.\ldots,l.
\end{equation}
which have a common modulus $x_l>0$  (a center of the string), a parity $v_l=\pm
1$ and differ by multiples of $q_{l}$.

Now we are ready to formulate the {\it string hypothesis} ---  a central concept of this
analysis:
\begin{itemize}
\item 
 At large $Q$
each solution of the BA consists of strings. 
\item
Each solution can be labeled by spins
$\{l_j,l_{j-1},\ldots\}$ and parities
$\{v_{j},v_{j-1}, \ldots\}$ of strings, such that the total number of roots 
$\sum_{i=1}^k (2l_i+1)=Q-1$. We refer to the set of lengths and parities of strings
constituting the solution for a given energy level as to a string content of this level.
Not more than two  strings with  a
given length and parity may be found in a string content of the solution.
\item
The length of the longest string in a
string content of a given energy level is the Hall conductance of the corresponding
band: $2l+1=|\sigma (k)|$. The period of this string $q_l=e^{i\pi\eta_l}$ is uniquely
determined by the requirement that $\eta_l=\frac{P_l}{2l+1}$ is the best approximant for the
period $\eta$, so that $q_l^{2l+1}=\pm 1$. 
\item
The parity of the longest string is
$v_l = -iq_l^{l+1/2}\nu=(-1)^{\left [\eta l\right]}\nu$.
The center of the longest string is $x_l=1+{\cal
O}(1/l)$.
\item
The remaining roots of the state is a solution of the BA equation for the parent state of
the parent generation.
\end{itemize}
The string hypothesis states that
\begin{equation}
\Psi^{\rm daugther}(z)\approx\prod_{m=-l}^{l}(z-x_lv_l q_{l}^m)\Psi^{\rm parent}(z).
\label{PsiAnsatz}
\end{equation}
and that $2l+1=|\sigma(k)|$ is the absolute value of the Hall conductance of the "daughter"
band.

This simple hypothesis allows one to construct a complete set of wave functions
 by virtue of the {\it iterative procedure}:
 
Starting from an irrational $\eta$, we first generate a hierarchical tree. Let us
choose a branch of the tree $J$.  Find
the Hall conductance of the bands belong to the branch down to the origin. This
determines the lengths,
periods and parities of the string content and therefore  zeros of the wave function of the
chosen branch. 

The only unknowns are the centers of strings. They, however approach
1 with accuracy ${\cal O}(1/l)$. The accuracy of the recursive eq.(\ref{PsiAnsatz}) is
${\cal O}(l^{-2})$. The string content of a state (i.e., lengths and parities of strings) is 
a topological characteristics, while the centers of strings are not.

The strings hierarchy has been obtained by analysis of singularities of BA (see
\cite{ATW97}). As it was expected 
 a set of possible lengths of strings is a 
 set of Takahashi-Suzuki numbers, known in the Bethe- Ansatz literature.  Eq.(\ref{DE})
provides a relation between them and Hall conductances.

To illustrate the iterative procedure let us consider the bottom edges of the lowest band of the spectrum
and choose
$\eta=\frac{\sqrt{5}-1}{2}$ to be the golden mean. 
The sequence of rational approximants is given by ratios of subsequent Fibonacci numbers $\eta_i=
\frac{F_{i-1}}{F_i}$, where the $F_i$
are Fibonacci numbers
($F_i=F_{i-2}+F_{i-1}$ and
$F_0=F_1=1$). The set of Hall conductances = Takahashi-Suzuki numbers = allowed lengths of strings are again Fibonacci numbers: $Q_{k-1} = F_{k-1}$.
 The considered branch of hierarchical tree connects edges
($\tau=-1,\kappa=1,\mu=1$) of the lowest bands of generations
$\eta_{3k}=F_{3k-1}/F_{3k}$. Their string content consists of pairs of strings
with length $2l_n+1=F_{3n+1}$, $n=0,1,\ldots,k-1$, parities $+1$ and inverted
centers $x_k$ and $x_k^{-1}$. According to the string hypothesis the wave function
of this state is
\begin{eqnarray}
\label{emp-wf-bottom}
\Psi(z|\eta_{3k}) \approx \prod_{n=0}^{k-1} \prod_{j=-l_n}^{l_n} 
(z-x_{l_n}q_{l_n}^j)(z-x_{l_n}^{-1}q_{l_n}^j). \end{eqnarray}
Centers of the strings $x_{l_n}$ are close to 1 but can not be obtained 
from the string hypothesis alone.

5. {\it Gaps.}
A direct application of the string hypothesis, suggested by J. Bellissard, is the calculation
of the gap distribution
$\rho(D)$, i.e., the number of gaps with magnitude between $D$ and $D+dD$. The result is 
 $\rho({\cal D})\sim
{\cal D}^{-3/2}$, which  essentially means that the width of the smallest gaps scales as
$D_{\rm min}\sim 1/Q^2$. This result confirms numerical analysis of
Ref.\cite{GKP91}.

6. {\it Scaling hypothesis and finite size corrections: stating the problem.} The string
hypothesis  solves the Bethe Ansatz equations with an accuracy
${\cal O}(Q^{-2})$, i.e., is asymptotically exact in the incommensurate limit
$Q\rightarrow\infty$. It alone
 gives
the explicit asymptotically exact form of  wave functions and provides the
hierarchical tree and topology of the Cantor set  spectrum.  However, the
most interesting quantitative characteristics of the spectrum are hidden in the
finite size corrections of the order of
$Q^{-2}$ to the bare value of strings. Among them are the anomalous dimensions of
the spectrum $\epsilon_{\cal J}$. They
depend on the branch  and on arithmetics of $\eta$
(according to ref.\cite{HirKohm89}  exponents $\epsilon_J$ vary between $0.171$
 and $-0.374$ for  the golden mean
$\eta=\frac{\sqrt 5-1}{2}$).  Can anomalous dimensions be found analytically, by finding
finite size corrections to the string solutions? This is a technically involved but a
fascinating and important problem. Its solution would provide an ultimate information of the
spectrum and most interesting physical properties of the system. It also may suggest the
conformal bootstrap and operator algebra approach, which has been proven to be effective for
finding the finite size corrections of integrable systems, without the actual solving the
Bethe Ansatz.

7. I would like to acknowledge  inspiring numerics provided by Y. Hatsugai during
the initial stages of formulated the string hypothesis, and useful discussions with Y. Avron,
G. Huber,  M. Kohmoto,  R. Seiler and A. Zabrodin. I also would like to thank M. Ninomiya for
the hospitality I received in YITP, where the text of this lecture was written.

\end{document}